\documentstyle[aps,twocolumn,epsfig]{revtex}

\begin{document}
\draft \small \twocolumn[\hsize\textwidth\columnwidth\hsize\csname
@twocolumnfalse\endcsname \title{Two-mode theory of vortex stability in 
multicomponent Bose-Einstein condensates}

\author{V\'{\i}ctor M. P\'erez-Garc\'{\i}a and Juan J. Garc\'{\i}a-Ripoll}

\address{ Departamento de Matem\'aticas, Escuela
 T\'ecnica Superior de Ingenieros Industriales\\ Universidad de
Castilla-La Mancha, 13071 Ciudad Real, Spain } \date{\today}

\maketitle


\begin{abstract}
 We study the stability and dynamics of vortices in two-species
condensates using a two mode model. The recent experimental results obtained
at JILA (M. R. Matthews, {\em et al.}, Phys. Rev. Lett. 83 (1999) 2498) and theoretical predictions based on the Gross-Pitaevskii equations are verified. We also make
an exhaustive analysis of the stability properties of the system when the relative populations of the two species and/or their relative scattering lengths are changed and prove that stabilization of otherwise unstable configurations can be attained by controlling the relative population of both species.
\end{abstract}

 ]


\narrowtext


\section{Introduction}

 Vortices appear in many different physical contexts ranging from
classical phenomena such as fluid mechanics \cite{Fluids} and
nonlinear Optics \cite{Kivshar} to purely quantum phenomena such as
superconductivity \cite{supercond} and superfluidity \cite{Superflu}.

 In fact a vortex is the simplest topological defect one can
construct \cite{Mermim} : in a closed path around a vortex the phase of the involved
field undergoes a $2\pi$ winding and stabilizes a zero value of the
field placed in what is called the vortex core.  The vortex is usually
stabilized by topological constraints since removing the phase
singularity implies an effect on the boundaries of the system which is
difficult to achieve using only local perturbations.

 In particular, the concept of a vortex is central to our
understanding of superfluidity and quantized flow. This is why after
the experimental realization of Bose-Einstein condensates (BEC) with
ultracold atomic gases \cite{experim} the question of whether atomic
BEC's are superfluids has triggered the analysis of vortices. The main
goals have been to propose a robust mechanism to generate vortices
\cite{generate} and detect them \cite{detect}. Finally another research area
has been the analysis of vortex stability \cite{stability,Rokhsar,Vortices,Fetter2}.

 Although most of the theoretical effort has concentrated on single
condensate systems, the final experimental realization of BEC vortices
\cite{MMatt99} following the proposal of Ref. \cite{Williams99} was attained
with a two-species Rb condensate. The two species correspond to two 
different hyperfine levels of Rb, denoted $|1\rangle$ and $|2\rangle$.
 In Ref. \cite{MMatt99} it was found that the vortex is
stable in only one of the two possible configurations. The stable configuration corresponds
to the vortex placed on the $|1\rangle$ state, which is the one with the larger scattering length. The other possibility, i. e. when the vortex is placed in the $|2\rangle$ state, which is the one with the lowest self-interaction coefficient,
leads to some kind of instability.  To simplify the notation in what follows we will 
use a shorthand notation for these states calling $|1,0\rangle$ to the state where the vortex is
put in $|1\rangle$ and $|0,1\rangle$ to the state where the vortex is placed in $|2\rangle$.

In  a recent work \cite{PRL3} we have shown, using numerical simulations, that the origin of the instability of state $|0,1\rangle$ is purely dynamical and can be understood within the framework of mean field theories for the double condensate system. A consequence of that analysis is to clarify the instability mechanism which does not lead to  expulsion of the vortex from condensate, but to the stablishment of a complex state where the phase singularity is periodically trasfered from one specie to the other.

In this paper our intention is to understand in more detail the instability mechanism and to
study configurations which can be obtained experimentally by varying the relative populations of the two species. The results are also applicable
to other multiple condensate systems where the different scattering lenghts have  values different from those of Rubidium. To make the analysis simpler and to ease the physical interpretation we make most of the analysis using a simplified two-mode model whose predictions 
are in close agreement with the experiment of Ref. \cite{MMatt99} and the numerics of Ref. \cite{PRL3} as well as with numerical simulations of the full three-dimensional mean field equations ruling the phenomenon.

Our plan is as follows: In Sec. \ref{themodel} we  present our problem in appropriate form and obtain the simplified equations for the two-mode model. In Sec. \ref{ddays} we apply the 
two-mode model to the situation discussed in the previous analysis \cite{MMatt99,PRL3}. In Sec. \ref{stability} we make a complete stability analysis of the relevant configurations and make some predictions which are experimentally testable. Finally in Sec. \ref{conclu} we summarize our conclusions.

\section{The model}
\label{themodel}

\subsection{Mean field equations for the two-condensate system}

 In this work we will use the zero temperature
approximation, in which collisions between the condensed and non
condensed atomic clouds are neglected. In the two species case this
leads to a pair of coupled Gross-Pitaevskii equations (GPE) for the condensate wavefunctions
of each specie
\begin{mathletters}
\label{GPE}
\begin{eqnarray}
i \hbar \frac{\partial}{\partial t} \Psi_1 &=& \left[ - \frac{\hbar^2\nabla^2}{2m} + 
 V_1 + U_{11} |\Psi_1|^2 + U_{12} |\Psi_2|^2 
 \right] \Psi_1, \\
i \hbar \frac{\partial}{\partial t} \Psi_2 &=& \left[ - \frac{\hbar^2\nabla^2}{2m} +
  V_2 + U_{21} |\Psi_1|^2 + U_{22} |\Psi_2|^2 
 \right] \Psi_2.
\end{eqnarray}
\end{mathletters}
Where $U_{ij} = 4\pi \hbar^2 a_{ij}/m$, $a_{ij}$ being the $s$-wave scattering lenghts of $1-1$, $(a_{11})$,
$2-2$ $(a_{22})$,  and $1-2$ $(a_{12})$ binary collisions.

To simplify the formalism we assume
that both potentials are of the form $V_1(\vec{r}) = V_2(\vec{r}) = \frac{1}{2}m
\omega^2 r^2.$

Next we change to a new set of units based on the trap characteristic
length, $a_0=\sqrt{\hbar/m\omega}$, and period, $\tau=1/\omega$
defined as $x \rightarrow x/a_0, t \rightarrow t/\tau$, $u_{ij} = 4\pi
N_j a_{ij}/a_0$ and $\Psi_j({\bf x}) = N_j \psi_j({\bf x})$.
Equations (\ref{GPE}) conserve the number of particles on each hiperfine level
so that we may choose
\begin{equation}
\int|\psi_1(\vec{r})|^2= \int|\psi_2(\vec{r})|^2 \equiv  1.
\end{equation}
This choice implies that the  particle number of each specie appears on the nonlinear coefficients $u_{ij}$.

The experimental results \cite{MMatt99} and our previous theoretical analysis \cite{PRL3} 
correspond to systems in which the number of particles is the same for each component, $N_1=N_2=N$, 
but in general one could allow any proportion between the populations of the different levels.

With the previous reescaling, the GPE for the multicomponent system read
\begin{mathletters}
\label{simple}
\begin{eqnarray}
i  \frac{\partial}{\partial t} \psi_1 &=& \left[ - \frac{1}{2} + 
 \frac{1}{2}r^2 + u_{11} |\psi_1|^2 + u_{12} |\psi_2|^2 
 \right] \psi_1, \\
i  \frac{\partial}{\partial t} \psi_2 &=& \left[ - \frac{1}{2} +
  \frac{1}{2}r^2 + u_{21} |\psi_1|^2 + u_{22} |\psi_2|^2 
 \right] \psi_2.
\end{eqnarray}
\end{mathletters}

Since the realistic values of Rb scattering lengths are in the proportion $a_{11}:a_{12}:a_{22} = 1.00:0.97:0.94$
\cite{Scattering-lengths}, the coefficients of the matrix of nonlinear coefficients satisfy the relations $u_{11}/u_{12} = a_{11}N_1/a_{12}N_2, u_{21}/u_{22} = a_{12}N_1/a_{22}N_2$, which means that except for the particular case in which $N_1 = N_2 = N,$ this matrix is nonsymmetric.
In terms of the population imbalace $\beta = N_2/N_1,$ and for a fixed total $N$ the matrix can be written as
\begin{equation}
\left(\begin{array}{cc} u_{11} & u_{12} \\ u_{21} & u_{22} \end{array} \right)
= \frac{4\pi a_{11} N}{a_0} \left(\begin{array}{cc} \frac{1.00}{1+\beta} & \frac{0.97\beta}{1+\beta}  \\ \frac{0.97}{1+\beta} & \frac{0.94\beta}{1+\beta}  \end{array} 
\right)
\end{equation}

\subsection{Derivation of a two-mode model}

 In our previous work \cite{PRL3} 
we considered the full problem and worked on the basis of the full GPE to prove
 the instability of the stationary solution $|0,1\rangle$, as well as the stability of state $|1,0\rangle$ for typical experimental conditions. The stability analysis based on the full GPE demonstrated that the instability was mediated by the growth of a core mode. This fact makes plausible the description of the two-condensate dynamics by the use of only two modes for each level one corresponding to the vortex and other to the core mode. This approach, which corresponds to retaining the stationary plus active modes and has been used succesully in the analysis of other nonlinear problems \cite{oscar}, should work at least in the linear regime where the perturbations are small. Mathematically, the idea is to approximate
\begin{mathletters}
\label{formula}
\begin{eqnarray}
\psi_1({\bf x}) \simeq a(t) \psi_{g1}({\bf x}) + b(t) \psi_{e1}({\bf x}), \\
\psi_2({\bf x}) \simeq c(t) \psi_{g2}({\bf x}) + d(t) \psi_{e2}({\bf x}).
\end{eqnarray}
\end{mathletters}
where $\psi_{gj}(x)$ is the spatial wavefunction of the ground state or core mode
for species $|j\rangle$ and $\psi_{ej}(x)$ corresponds to a representation of the single vortex  wavefunction.
This approximation implies some loss of information on the dynamics but is not essential for our results as will be shown later. We will see that this ansatz reflects the essentials of the dynamics with good accuracy. 

 $\psi_g, \psi_e$ can be choosen as any approximation to the 
ground and first excited states of the single specie equations, provided they are orthogonal. 
 Our choice will be to use the eigenfunctions of the $d-$dimensional harmonic oscillator which are the exact solutions in the linear case and allow simple manipulation since their analytic form is known:
\begin{mathletters}
\begin{eqnarray}
\psi_g({\bf x}) & = & \left(\frac{1}{\pi}\right)^{d/2} e^{-r^2/2}, \\
\psi_e({\bf x}) & = & \left(\frac{2}{d\pi}\right)^{d/2} r e^{-r^2/2} e^{i\theta}.
\end{eqnarray}
\end{mathletters}
Other choices are possible with the only change of several coefficients related to integrals involving $\psi_g$ and $\psi_e$.
In our treatment we will consider simultaneously the two and three dimensional 
configurations. 
It was shown in a previous work concerning single condensates \cite{Vortices},
 that the transition from a spherical trap to a 
pancake preserves the shape and number of unstable modes. This fact motivated us in Ref. \cite{PRL3} to use a simpler model in which the
dependency on the axial direction has been integrated out. Our present analysis 
applies equally to the simplified 2D situation used in Ref. \cite{PRL3}
as well as to the full 3D problem and proves that there are no essential differences between the two and three-dimensional models for the phenomena studied here.

 Inserting ansatz (\ref{formula}) into GP equation (\ref{simple}) and proyecting on $\psi_{gj}$ and $\psi_{ej}$ the following set of coupled nonlinear ODE are obtained
\begin{mathletters}
\label{twomode}
\begin{eqnarray}
i\dot{a} & = & -a E_g + u_{11} a \left(\gamma_1 |a|^2 + 2 \gamma_2 |b|^2\right) \nonumber \\
         &  & + u_{12} a \left( \gamma_1 |c|^2 + \gamma_2|d|^2 \right) + u_{12} \gamma_2 b c d^* ,\\
i\dot{b} & = & -b E_e + u_{11} b \left(\gamma_3 |b|^2 + 2 \gamma_2 |a|^2\right) \nonumber \\
         &   & +u_{12} b \left( \gamma_2 |c|^2 + \gamma_3|d|^2 \right) + u_{12} \gamma_2 a c^* d ,\\
i\dot{c} & = & -c E_g + u_{22} c \left(\gamma_1 |c|^2 + 2 \gamma_2 |d|^2\right) \nonumber \\
         &  & +u_{21} c \left( \gamma_1 |c|^2 + \gamma_2|d|^2 \right) + u_{21} \gamma_2 d a b^* ,\\
i\dot{d} & = & -d E_e + u_{22} d \left(\gamma_3 |d|^2 + 2 \gamma_2 |c|^2\right) \nonumber \\
         &  & + u_{21} d \left( \gamma_2 |a|^2 + \gamma_3|b|^2 \right) + u_{21} \gamma_2 c a^* b ,
\end{eqnarray}
\end{mathletters}
where $\gamma_1 = (\psi_g^2 , \psi_g^2), \gamma_2 = (\psi_g^2, \psi_e^2), \gamma_3 = (\psi_e^2, \psi_e^2)$ and $E_g = (\psi_g, H_0 \psi_g), E_e = (\psi_e, H_0 \psi_e)$ being $H_0 = -\frac{1}{2} \triangle + \frac{1}{2} r^2$. For our particular choice of $\psi_g$ and $\psi_e$ the numerical values of these coefficients are $\gamma_1^{2d}= 1/2\pi, \gamma_2^{2d} = 1/4\pi$, $\gamma_3^{2d} = 
1/4\pi, E_g^{2D} = 1, E_e^{2D} = 2$, $\gamma_1^{3d}= 1/(2\pi)^{3/2}, \gamma_2^{3d} = 1/2(2\pi)^{3/2}$, $\gamma_3^{3d} = 5/12(2\pi)^{3/2}$, $E^{3D}_g = 3/2, E^{3D}_e = 5/2$.

Equations (\ref{simple}) satisfy discrete conservation laws corresponding to the number of particles of each species and angular momentum
\begin{mathletters}
\begin{eqnarray}
|a|^2 + |b|^2 & = & 1, \\
|c|^2 + |d|^2 & = & 1, \\
|b|^2 + |d|^2 & = & L_0, 
\end{eqnarray}
\end{mathletters}
$L_0$ being the angular momentum of the initial data. There is another conservation law
for energy which is not relevant for our purpouses. 
 
 It is convenient to change to the modulus-phase representation as, $a =\rho_a e^{i\phi_a},
b =\rho_b e^{i\phi_b}, c =\rho_c e^{i\phi_c}, d =\rho_d e^{i\phi_d}$. Eqs. (\ref{twomode}) 
become
\begin{mathletters}
\label{complete}
\begin{eqnarray}
\dot{\rho}_a & = & u_{12} \gamma_2 \rho_b \rho_c \rho_d \sin \left(\phi_b + \phi_c- \phi_d -\phi_a \right),\\
\dot{\phi}_a & = & E_g  - u_{11}\left(\gamma_1 \rho_a^2 + 2 \gamma_2 \rho_b^2\right) - 
u_{12} \left(\gamma_1 \rho_c^2 + \gamma_2 \rho_d^2\right) \nonumber \\
&  & -u_{12} \gamma_2 \frac{\rho_b \rho_c \rho_d}{\rho_a} \cos \left(\phi_b + \phi_c-\phi_d -\phi_a\right), \\
\dot{\rho}_b & = & u_{12} \gamma_2 \rho_a \rho_c \rho_d \sin \left(\phi_a + \phi_d-\phi_a -\phi_c\right),\\
\dot{\phi}_b & = & E_e  - u_{11}\left(\gamma_3 \rho_b^2 + 2 \gamma_2 \rho_a^2\right) - 
u_{12} \left(\gamma_2 \rho_c^2 + \gamma_3 \rho_d^2\right) \nonumber \\
&  & - u_{12} \gamma_2 \frac{\rho_a \rho_c \rho_d}{\rho_b} \cos \left(\phi_d + \phi_a-\phi_c -\phi_b\right) ,\\
\dot{\rho}_c & = & u_{21} \gamma_2 \rho_b \rho_a \rho_d \sin \left(\phi_b + \phi_c-\phi_d -\phi_a\right), \\
\dot{\phi}_c & = & E_g  - u_{22} \left(\gamma_1 \rho_c^2 + 2 \gamma_2 \rho_d^2\right) - 
u_{21} \left(\gamma_1 \rho_a^2 + \gamma_2 \rho_b^2\right) \nonumber \\
&  & -u_{21} \gamma_2 \frac{\rho_b \rho_a \rho_d}{\rho_c} \cos \left(\phi_b + \phi_c-\phi_d -\phi_a\right), \\
\dot{\rho}_d & = & u_{21} \gamma_2 \rho_a \rho_b \rho_c \sin \left(\phi_d + \phi_a-\phi_a -\phi_c\right),\\
\dot{\phi}_d & = & E_e  - u_{22} \left(\gamma_3 \rho_d^2 + 2 \gamma_2 \rho_c^2\right) - 
u_{21} \left(\gamma_2 \rho_a^2 + \gamma_3 \rho_b^2\right) \nonumber \\
&  & - u_{21} \gamma_2 \frac{\rho_a \rho_c \rho_b}{\rho_d} \cos \left(\phi_d + \phi_a-\phi_c -\phi_b\right).
\end{eqnarray}
\end{mathletters}
Despite the apparent complexity of this system it can be seen that the four phase variables can be reduced to only one defining $\Phi = \phi_b + \phi_c - \phi_a - \phi_d,$
and then
\begin{mathletters}
\label{rho}
\begin{eqnarray}
\dot{\rho}_a & = & u_{12} \gamma_2 \rho_b \rho_c \rho_d \sin \Phi, \\
\dot{\rho}_b & = -&  u_{12} \gamma_2 \rho_a \rho_c \rho_d \sin \Phi,\\
\dot{\rho}_c & = -& u_{21} \gamma_2 \rho_b \rho_a \rho_d \sin \Phi,\\
\dot{\rho}_d & = & u_{21} \gamma_2 \rho_a \rho_b \rho_c \sin \Phi,\\
\dot{\Phi} & = & \gamma_a \rho_a^2 + \gamma_b \rho_b^2 + \gamma_c \rho_c^2 + \gamma_d \rho_d^2 \nonumber \\
&  & + \gamma_2 \left[ u_{12} \left(\frac{\rho_b \rho_c \rho_d}{\rho_a} - 
\frac{\rho_a \rho_c \rho_d}{\rho_b}\right) \right. \nonumber \\
  & & + u_{21} \left.
 \left(\frac{\rho_b \rho_c \rho_a}{\rho_d} - 
\frac{\rho_a \rho_b \rho_d}{\rho_c}\right) \right].
\end{eqnarray}
\end{mathletters}
We have now five equations in (\ref{rho}) plus four conservation laws, which means that the system can be (at least formally) integrated. This fact excludes the possibility of chaotic behavior in the system.

These equations can be further simplified by defining density variables related to $\rho_j^2$ 
and $X = \rho_a \rho_b \rho_c \rho_d$ and using the conservation laws. We will not follow this route since
all the simplified models such as the one presented in Eqs. (\ref{rho}) have singularities when any of the densities is 
zero. This fact makes the equations unuseful for the purpouse of stability analysis, since the stationary states
are singular solutions of these systems.

\section{Dynamics in the physically relevant case} 
\label{ddays}

\begin{figure}
\epsfig{file=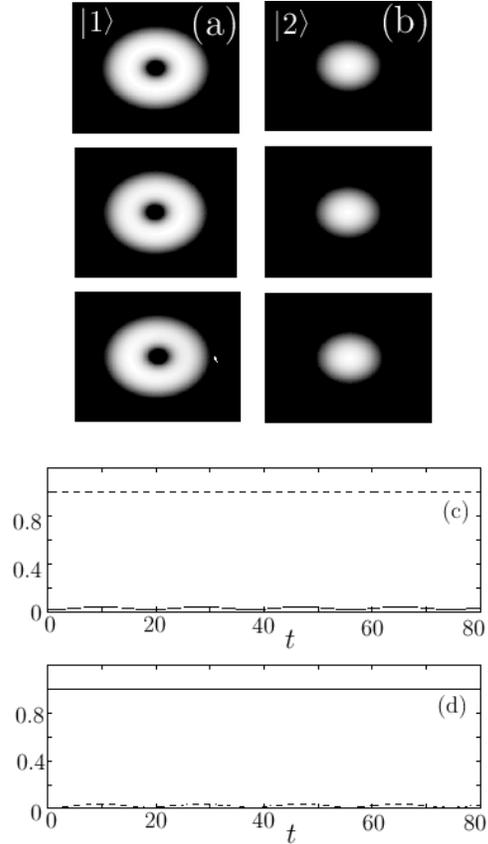,width=6.3cm}
\caption{Stability of the configuration $|1,0\rangle$. Snapshots of the spatial density of (a) $|1\rangle$ and (b) $|2\rangle$. 
Evolution the amplitudes of the modes with time (c) $|a|$ (solid line) and $|b|$ (dashed line);
 (d) $|c|$ (solid line) and $|d|$ (dashed line).}
\label{fig0}
\end{figure}

\begin{figure}
\epsfig{file=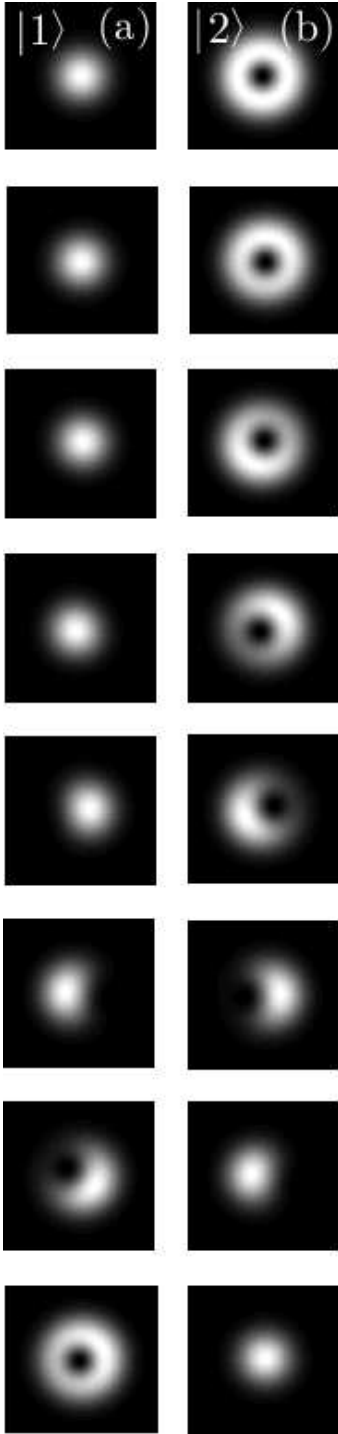,width=4.5cm}
\caption{ Snapshots of the evolution of an unstable vortex (state $|0,1\rangle$). 
 Evolution of the spatial density of (a)$|1\rangle$ and (b) $|2\rangle$.}
\label{figa}
\end{figure}

\begin{figure}
\epsfig{file=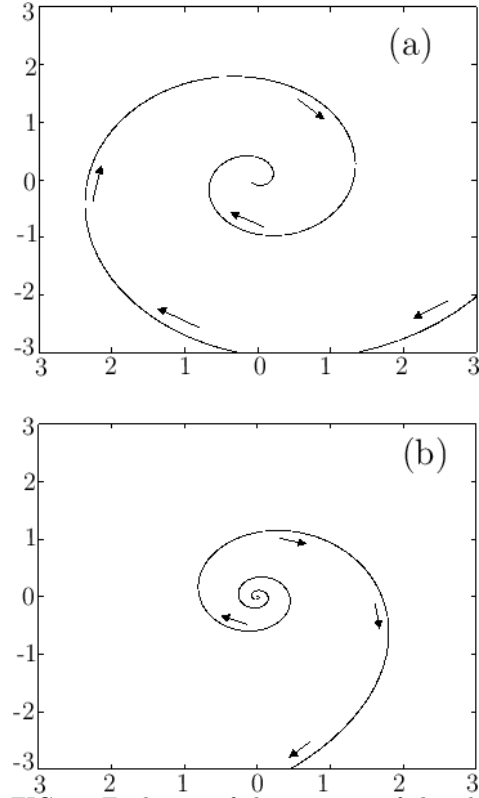,width=6.2cm}
\caption{Evolution of the position of the phase singularity corresponding 
to the simulation shown in Fig. \ref{figa}. (a) Phase singularity in $|1\rangle$  (b) Phase singularity in $|2\rangle$.} 
\label{figb}
\end{figure}

\begin{figure}
\epsfig{file=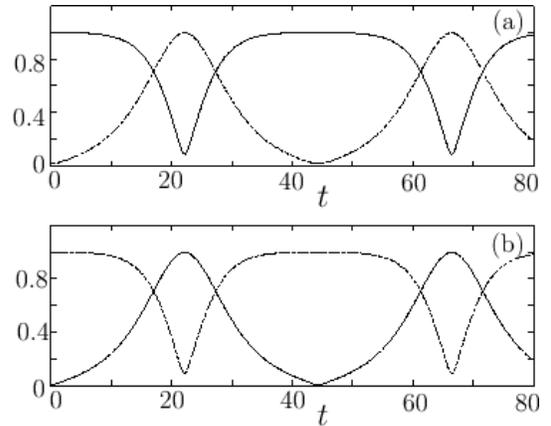,width=7.0cm}
\caption{Evolution of the amplitudes of the modes (a) $|a|$ (dashed line) and $|b|$ (solid line); (b) $|c|$ (dashed line) and $|d|$ (solid line). The simulation is done with a random perturbation to the configuration with vortex in $|2\rangle$, i. e., $a(0) = e^{i r_2} \sqrt{1-\epsilon_1^2}, b(0) = \epsilon_1 e^{i r_3}$,  $c(0) = \epsilon_2 e^{i r_4}; d(0) = \sqrt{1-\epsilon_2^2}$. $\epsilon_1$ and $\epsilon_2$ are random numbers
 uniformly distributed between 0 and 0.02. $r_j$ are random numbers uniformly distributed between 0 and 1.}
\label{figc}
\end{figure}

The configurations studied experimentally in
\cite{MMatt99} and numerically in \cite{PRL3} correspond to single vortices in $|1\rangle$ or $|2\rangle$ in the case where the populations of both hiperfine levels are equal. It means that 
$\beta =1$ and thus $u_{21} = u_{12}$. It is interesting to study the dynamics of small perturbations of the initial data $a(0) = 0, b(0) = 1, c(0) = 1, d(0) = 0$, which physically corresponds to the stationary state $|1,0\rangle$ and $a(0) = 1, b(0) = 0, c(0) = 0, d(0) = 1$ which corresponds to $|1,0\rangle$. These initial data correspond to periodic solutions of the amplitude equations (\ref{twomode}), which are
\begin{mathletters}
\begin{eqnarray}
a(t) = 0, b(t) = e^{i(E_e - u_{11}\gamma_3 -u_{12}\gamma_2)t}, \nonumber \\
c(t) = e^{i(E_g - u_{22}\gamma_1 -u_{21}\gamma_2)t}, d(t) = 0. \label{v1}\\
a(t) =  e^{i(E_e - u_{11}\gamma_1 -u_{12}\gamma_2)t}, b(t) = 0, \nonumber\\
 c(t) = 0, d(t) =  e^{i(E_g - u_{22}\gamma_3 -u_{21}\gamma_2)t}.\label{v2}
\end{eqnarray}
\end{mathletters}
respectively. 
 To have a clear picture of what is going on we have first simulated the dynamics of these 
states when small perturbations are added, the results being summarized in Figs. \ref{fig0} to \ref{figc}. 

It is clear in Fig. \ref{fig0} that the configuration which has a vortex in $|1\rangle$ is dynamically stable. However, 
when the vortex is placed in $|2\rangle$ an instability develops and the response to small perturbations is to trasfer the
vortex to $|1\rangle$ and start a periodic trasfer dynamics. The snapshots of the trasfer (evolution of the density) are shown
in Fig. \ref{figa}. In Fig. \ref{figb} it is seen how the phase singularity in $|2\rangle$ spirals out of the system while a 
phase singularity appears in $|1\rangle$ and occupies the center of the atomic cloud \cite{nota}. This dynamics is recurrent as can be seen from the evolution of the relevant variables (Fig. \ref{figc}).

\section{Stability theory}
\label{stability}

\subsection{Problem statement}

 Our numerical simulations of the reduced system (\ref{twomode}) show that in the equal population case, $N_1 = N_2 = N$, and for arbitrary
nonlinearities one of the possible stationary states of the system is stable while the other is unstable. It is our purpouse in this section to 
make a complete analysis of the stability of the system for any proportion of the populations $\beta = N_1/N_2$ and 
any value of the nonlinear coefficients (e.g. total number of particles, $N$, and scattering lengths $a_{ij}$). These results
could be specially relevant to predict for a specific multiple-condensate system the existence of stable vortex states. For the case 
of Rb the results can be applied to study the possibility of stabilizing different configurations.

\subsection{Stability of state $|1,0\rangle$}

When a vortex is placed at $|1\rangle$ the resulting stationary state is a periodic orbit described by Eq. (\ref{v1}). Its direct stability analysis using Eqs. (\ref{twomode}) would lead to time-dependent perturbation equations which should be analyzed using Floquet theory.
A way to circunvent partially this situation is to change to the rotating system defined as
\begin{mathletters}
\begin{eqnarray}
\tilde{a} & =& a e^{-i(E_g-u_{22}\gamma_1 - u_{21} \gamma_2)t}, \\
\tilde{b} & = & b e^{-i(E_e-u_{11}\gamma_3 - u_{12} \gamma_2)t}, \\
\tilde{c} & =& c e^{-i(E_g-u_{22}\gamma_1 - u_{21} \gamma_2)t}, \\
\tilde{d} & = & d e^{-i(E_e-u_{11}\gamma_3 - u_{12} \gamma_2)t}.
\end{eqnarray}
\end{mathletters}
Using these new variables the equations are
\begin{mathletters}
\label{22mode}
\begin{eqnarray}
\dot{\tilde{a}} & = & i \left(\gamma_1u_{22}+\gamma_2u_{21}\right) \tilde{a} - i u_{11} \tilde{a} \left(\gamma_1 |\tilde{a}|^2 + 2 \gamma_2 |\tilde{b}|^2\right) \nonumber \\
         &  & - i u_{12} \tilde{a} \left( \gamma_1 |\tilde{c}|^2 + \gamma_2|\tilde{d}|^2 \right) -i u_{12} \gamma_2 \tilde{b} \tilde{c} \tilde{d}^* ,\\
\dot{\tilde{b}} & = &  i \left(\gamma_3u_{11}+\gamma_2u_{12}\right) \tilde{b} -i u_{11} \tilde{b} \left(\gamma_3 |\tilde{b}|^2 + 2 \gamma_2 |\tilde{a}|^2\right) \nonumber \\
      &   & - i u_{12} \tilde{b} \left( \gamma_2 |\tilde{c}|^2 + \gamma_3|\tilde{d}|^2 \right) -i u_{12} \gamma_2 \tilde{a} \tilde{c}^* \tilde{d} ,\\
\dot{\tilde{c}} & = &  i \left(\gamma_1u_{22}+\gamma_2u_{21}\right) \tilde{c}  - i u_{22} \tilde{c} \left(\gamma_1 |\tilde{c}|^2 + 2 \gamma_2 |\tilde{d}|^2\right) \nonumber \\
         &  & - i u_{21} \tilde{c} \left( \gamma_1 |\tilde{c}|^2 + \gamma_2|\tilde{d}|^2 \right) -i u_{21} \gamma_2 \tilde{d} \tilde{a} \tilde{b}^* ,\\
\dot{\tilde{d}} & = & i \left(\gamma_3u_{11}+\gamma_2u_{21}\right) \tilde{d} - i u_{22} \tilde{d} \left(\gamma_3 |\tilde{d}|^2 + 2 \gamma_2 |\tilde{c}|^2\right) \nonumber \\
         &  & - i u_{21} \tilde{d} \left( \gamma_2 |\tilde{a}|^2 + \gamma_3|\tilde{b}|^2 \right) -i u_{21} \gamma_2 \tilde{c} \tilde{a}^* \tilde{b},
\end{eqnarray}
\end{mathletters}
and the stationary solution is an equilibrium point of Eqs. (\ref{22mode}): $\tilde{a}_0 = 0,
\tilde{b}_0 = 1, \tilde{c}_0 = 1, \tilde{d}_0 = 0$.
 To study its stability we linearize Eqs (\ref{22mode}) around the equilibrium point and define the perturbations through
\begin{mathletters}
\begin{eqnarray}
\tilde{a} & = & \tilde{a}_0 + \delta_a(t), \\
\tilde{b} & = & \tilde{b}_0 + \delta_b(t), \\
\tilde{c} & = & \tilde{c}_0 + \delta_c(t), \\
\tilde{d} & = & \tilde{d}_0 + \delta_d(t).
\end{eqnarray}
\end{mathletters}
Their evolution laws are
\begin{mathletters}
\begin{eqnarray}
\dot{\delta}_a & =& i \Delta_a \delta_a - i u_{12}\gamma_2 \delta_d^*, \\
\dot{\delta}_b & =& 0, \\
\dot{\delta}_c & =& 0, \\
\dot{\delta}_d & =& i \Delta_d \delta_d - i u_{21}\gamma_2 \delta_a^*, 
\end{eqnarray}
\end{mathletters}
where $\Delta_a = u_{22}\gamma_1+u_{21}\gamma_2-2\gamma_2u_{11}-\gamma_1u_{12}$, $\Delta_d = u_{11}\gamma_3+u_{12}\gamma_2-2u_{22}\gamma_2 - u_{21}\gamma_3$.
The perturbations for $\tilde{b}$ and $\tilde{c}$ have a neutral behavior and their evolution is ruled by quadratic terms. If we write the equations for
the perturbations and their complex conjugates to obtain the full stability spectrum we have
\begin{equation}
\frac{d}{dt} \left( \begin{array}{c}  {\delta}_a \\ {\delta}_b^* \\ {\delta}_d \\  {\delta}_d^* \end{array} \right) = 
i\left(\begin{array}{cccc}  \Delta_a & 0 & 0 & -  u_{12}\gamma_2  \\
                             0   & -\Delta_a &  u_{12}\gamma_2 & 0 \\
                             0 &  - u_{21}\gamma_2 & \Delta_d & 0 \\
                             u_{21} \gamma_2 & 0 & 0 & - \Delta_d \end{array}\right) \left( \begin{array}{c}  {\delta}_a \\ {\delta}_b^* \\  {\delta}_d \\ {\delta}_d^* \end{array} \right)
\end{equation}
The eigenvalues of this matrix can be obtained analytically, the result being
\begin{mathletters}
\begin{eqnarray}
\lambda_{1,2} & =& i \left[ \frac{\Delta_a - \Delta_d}{2} \pm \frac{1}{2} \sqrt{\left(\Delta_a+\Delta_d\right)^2-4 \gamma_2^2 u_{21}u_{12}}\right], \\
\lambda_{3,4} & =& i\left[ \frac{\Delta_a + \Delta_d}{2} \pm \frac{1}{2} \sqrt{\left(\Delta_a+\Delta_d\right)^2-4 \gamma_2^2 u_{21}u_{12}}\right].
\end{eqnarray}
\end{mathletters}
There is only one stability condition which is $|\Delta_a + \Delta_d| > 2 \gamma_2 \sqrt{u_{21}u_{12}}$. Since the $\Delta_j$ are functions of $\gamma_j$, which depend on the dimensionality, we must distinguish now results for the 2D and 3D cases separately.  In the two dimensional case we obtain
\begin{equation}
\label{2Dmode}
u_{11} + u_{12} > 2 \sqrt{u_{21}u_{12}},
\end{equation}
while for 3D the condition is
\begin{equation}
\label{3Dmode}
\frac{7}{6}u_{11} - \frac{1}{6} u_{21} + u_{12}> 2 \sqrt{u_{21}u_{12}}.
\end{equation}
Taking into account the fact that the numerical values of $a_{11}$ and $a_{12}$ are very close we find that Eqs. (\ref{2Dmode}) and (\ref{3Dmode}) are very simmilar. This is why we will use one of them, Eq. (\ref{2Dmode}), for the subsequent analysis.
If we write these equations in terms of $\beta$ and use the scattering length values of Rb, we obtain
the following stability condition
\begin{equation}
\label{ine1}
 \frac{a_{11}}{a_{12}}  + \beta  >  2 \sqrt{\beta}.
\end{equation}

 For Rb values the inequality  (\ref{ine1}) is allways satisfied, which proves that the configuration with a vortex in $|1\rangle$ is always linearly stable no matter what is the relation between the populations $\beta$, which implies also the stability of
the experimental configuration with vortex in $|1\rangle$ studied in Refs. \cite{MMatt99,PRL3}. 

It is also interesting that the stability properties do not depend on the total number of
particles but only on the relation between the populations \cite{coment}. 

\subsection{Stability of the vortex in $|2\rangle$ state}

\begin{figure}
\epsfig{file=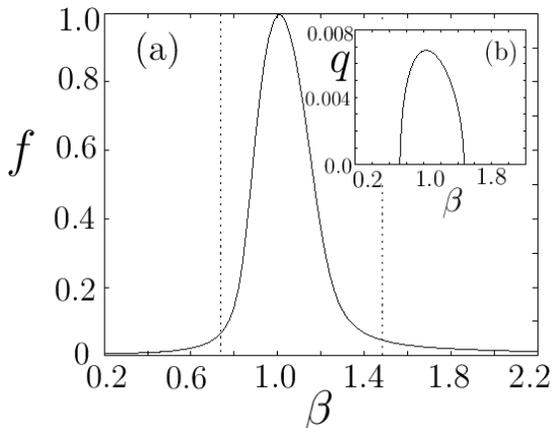,width=7.2cm}
\caption{Angular momentum trasfer as a function of $\beta$. All the angular momentum is put initially at 
component $|2\rangle$. (a) We simulate the dynamics and plot the maximum fraction of angular momentum at component $|1\rangle$, $f$, which is a measure of the instability, as a function of $\beta$. The vertical lines mark the points where the stability analysis predicts instability 
of the configurations. (b) Value of the real part of the eigenvalues leading to instability $\lambda = 4\pi N a_{11} q/a_{0}$ (c) Results of realistic numerical simulations of the fraction of angular momentum trasfered as a function of .} 
\label{trasfer2}
\end{figure}

The stability analysis of configuration $|0,1\rangle$, which corresponds to initial data $a_0 = 1, b_0 = 0, c_0 = 0, d_0 =1$, is completely equivalent to the previous one. In fact, elementary symmetry arguments allow to ensure that the 
result should be equivalent with the only interchange of the indices $1\leftrightarrow 2$, i.e., the stability condition is
\begin{equation}
 \frac{a_{22}}{a_{21}}\beta  + 1   >  2 \sqrt{\beta} \label{inem}\\
\end{equation}
This inequality is not verified out for a range of $\beta$ values. Solving the algebraic equation for $\beta$ one finds for the critical values
\begin{equation}
\frac{1}{\beta_c}  = \left(\frac{N_1}{N_2}\right)_c = 2 - \frac{a_{22}}{a_{12}} \pm \sqrt{1-\frac{a_{22}}{a_{21}}}.
\end{equation}
For the case of Rb it is found that the unstable range is $\beta \in [0.73,1.49]$. This result is interesting since it means that 
there are choices of the population imbalance $\beta$ which allow stabilization of the vortex in $|2\rangle$. We have analyzed the ratio of angular momentum trasfer from component $|2\rangle$ to $|1\rangle$ as a function of $\beta$ from numerical simulations of Eqs. (\ref{twomode}).
 The results are presented in Fig. \ref{trasfer2}. This is one of the main results of the paper and a prediction which can be experimentally tested. We have also made compared these results with numerical simulations of the Gross-Pitaevskii equation with good qualitative agreement. Of course {\em the exact range depends on the size of the perturbation added to destabilize the vortex}, this means that Fig. 5 must be taken only as orientative since the exact profile can change when the size of the perturbations added are changed. However the $\beta$ stability ranges found by numerical simulations for different perturbations are simmilar to the prediction of the two mode theory.

\subsection{Nonlinear stability analysis}

 The nonlinear stability analysis is quite technical and difficult for this particular case and will be the subject of future work. However, there are indications that the stable configuration $|1,0\rangle$ should be sensitive to appropriate small finite amplitude perturbations. The reason is very simple, if one looks to Fig. 4(b) when the vortex is almost completely trasfered to state $|1\rangle$ (for a time near 20 time units) we are in a configuration which is close to the stable state $|1,0\rangle$, but which is unstable. 

 In fact we have added finite amplitude perturbations to the configuration $|1,0\rangle$ and find that a periodic trafer dynamics is also induced very simmilar to the dynamics of $|2,0\rangle$. The main difference is that $|1,0\rangle$ is linearly stable, which makes this configuration more robust but yet not completely stable.

\section{Conclusions and discussion}
\label{conclu}

 Summarizing the work presented in this paper we have completed the task of analyzing stability properties of vortices in double Rb condensates, although our theory is much more general and can be applied to any two (multiple)-condensate system. Our treatment is based on a simplified two-mode model which captures many of the relevant dynamical features of the problem. We have attained several goals in this work. First, the stability results of Refs. \cite{MMatt99,PRL3} are reproduced for the $N_1 = N_2$ case. Second, the instability mechanism consisting in vortex exchange between the two species is supported and
described in detail here. Third, we rise a new prediction which consists on the fact that population imbalances can stabilize vortices in $|2\rangle$ states and also prove that vortices in $|1\rangle$ can be destabilized by adding finite amplitude perturbations to the initial data. These predictions can be tested with current experimental setups and can be another tests of the existence of purely 
dynamical instabilites in the two-condensate system of Ref. \cite{MMatt99}. 

 The proposed possibility of making the condensate in $|2\rangle$ stable or unstable by controlling the population ratio is interesting from the viewpoint of condensate engineering.
We hope that this works helps in the task of understanding the complex dynamics of vortices 
in Bose-Einstein condensates.

\acknowledgements

This work has been partially supported by the DGICYT under grant PB96-0534.



\begin{references}
\bibitem{nota}{The evolution of the phase singularities can be obtained from the condition $\hbox{Real}(\psi_j) = \hbox{Imag}(\psi_j) = 0$. The position of the phase singularity at $|2\rangle$ is given by the equations: $x_s(t) = |a(t)| \sin \left( \phi_a(t) - \phi(b) + \pi\right)/|b(t)|$, $y_s(t) = |a(t)| \cos \left( \phi_a(t) - \phi(b) + \pi\right)/|b(t)|$. To obtain the trayectory of the singularity at $|2\rangle$ can be obtained by changing $a$ and $b$ by $c$ and $d$ respectively.}
\bibitem{coment}{The stability properties also depend essentially on the scattering lengths, which in our case are fixed since we are dealing with specific hyperfine levels of Rb atoms.}
\end{references}
\end{document}